\documentclass[pdftex,iop,apjl]{emulateapj} % Linewraps are better with apjl option

\usepackage{longtable}
\usepackage{epstopdf}  % allows you to use the pdflatex script from JCV
\usepackage{ifthen}
\usepackage{rotating}
\usepackage{subfigure} % allows numbering figures 1a, 1b, etc.
\usepackage{booktabs}

\usepackage{appendix}
\usepackage{color}

\submitted{Accepted for publication in {\it The Astrophysical Journal Letters}, 2016 April 23}
\shorttitle{{\sc Comet 322P/SOHO 1: An asteroid with the smallest perihelion distance?}}
\shortauthors{{\sc Knight et al.}}

\def\deg{$^\circ$}

\def\icarus{Icarus}

\begin{document}
\bibliographystyle{apj}

\title{Comet 322P/SOHO 1: An asteroid with the smallest perihelion distance?\altaffilmark{*}}

\author{Matthew M. Knight\altaffilmark{1,2,3}, {Alan Fitzsimmons\altaffilmark{4}}, Michael S.~P. Kelley\altaffilmark{2}, Colin Snodgrass\altaffilmark{5}}

\altaffiltext{1}{Contacting author: mmk8a@astro.umd.edu.}
\altaffiltext{2}{University of Maryland, Department of Astronomy, 1113 Physical Sciences Complex, Building 415, College Park, MD 20742, USA}
\altaffiltext{3}{Observations obtained while at Lowell Observatory, 1400 W. Mars Hill Rd, Flagstaff, Arizona 86001, USA}
\altaffiltext{4}{Astrophysics Research Center, School of Mathematics and Physics, Queen’s University Belfast, Belfast BT7 1NN, UK}
\altaffiltext{5}{Planetary and Space Sciences, Department of Physical Sciences, The Open University, Milton Keynes, MK7 6AA, UK}
\altaffiltext{*}{Based on observations collected at the European Organisation for Astronomical Research in the Southern Hemisphere under ESO programme 095.C-0853, with Lowell Observatory's Discovery Channel Telescope, and with {\it Spitzer Space Telescope} under program 11104.}

\begin{abstract}
We observed comet 322P/SOHO 1 (P/1999 R1) from the ground and with the {\it Spitzer Space Telescope} when it was between 2.2 and 1.2 AU from the Sun. These are the first observations of any {\it SOHO}-discovered periodic comet by a non-solar observatory, and allow us to investigate its behavior under typical cometary circumstances. 322P appeared inactive in all images. Its lightcurve suggests a rotation period of $2.8{\pm}0.3$ hr and has an amplitude $\gtrsim$0.3 mag, implying a density of at least 1000 kg m$^{-3}$, considerably higher than that of any known comet. It has average colors of $g'-r' = 0.52{\pm}0.04$ and $r'-i' = 0.04{\pm}0.09$. We converted these to Johnson colors and found that the $V-R$ color is consistent with average cometary colors, but $R-I$ is somewhat bluer; these colors are most similar to V- and Q-type asteroids. Modeling of the optical and IR photometry suggests it has a diameter of 150--320 m and a geometric albedo of 0.09--0.42, with diameter and albedo inversely related. Our upper limits to any undetected coma are still consistent with a sublimation lifetime shorter than the typical dynamical lifetimes for Jupiter Family Comets. These results suggest that it may be of asteroidal origin and only active in the {\it SOHO} fields of view via processes different from the volatile-driven activity of traditional comets. If so, it has the smallest perihelion distance of any known asteroid. 
\end{abstract}


\section{INTRODUCTION}
Sungrazing orbits are predicted to be a major end-state of main-belt asteroids and near-Earth objects \citep{farinella94,gladman97} but are yet to be observed. Models of Solar System evolution predict the numbers and original source regions of small perihelion distance ($q$) objects and also expect such orbits to be common \citep{bottke02,greenstreet12}. Recent work by \citet{granvik16} has highlighted the scarcity of small-$q$ asteroids and identified possible mechanisms for their destruction. Of the 34 known asteroids with $q<0.15$ AU\footnote{{\url http://ssd.jpl.nasa.gov/sbdb\_query.cgi?}, retrieved March 2016}, the smallest has $q=0.071$ AU, which is well beyond typical ``sungrazing'' distances of $\sim$0.01 AU (e.g., \citealt{knight13b}). 

A possibly overlooked source of small-$q$ asteroids is the database of objects discovered by solar observatories, primarily {\it Solar and Heliospheric Observatory} ({\it SOHO}). More than 3000 comets have been discovered in {\it SOHO} images since 1996 (e.g., \citealt{biesecker02,knight10d,lamy13}). The vast majority belong to one of several families, whose members are dynamically related to each other and are apparently produced by cascading fragmentation from a single progenitor comet into numerous higher generation fragments.

Only $\sim$100 {\it SOHO}-discovered comets are not dynamically linked to the major near-Sun comet families. Most do not display a coma or tail, however, these ``sporadic'' objects are designated as comets since all prior objects seen at these distances have been of apparently cometary origin. This assumption is reasonable given that {\it SOHO}'s limiting magnitude would necessitate a bare asteroid being $\gtrsim10$ km in diameter to be observed; numerous such large objects are unlikely to be missed in the modern survey era, especially by NEOWISE where they would be particularly bright \citep{mainzer11}. Although these objects are almost certainly active when in the {\it SOHO} fields of view ($\lesssim$0.15 AU), this activity is not necessarily due to the traditional ``cometary'' mechanism of sublimation of volatile ices. \citet{jewitt10} and \citet{li13b} argue that (3200) Phaethon produces dust near perihelion (0.14 AU) via non-traditional means such as thermal fracture, while \citet{kimura02} and others have shown that silicates and other refractory materials begin to sublimate at even smaller distances. 

Of the sporadic near-Sun objects, comet 322P/SOHO~1 (henceforth 322P), is the most promising for study at larger distances to investigate whether or not it is of a traditional cometary origin. 322P was discovered in {\it SOHO} images in 1999 and originally designated C/1999~R1. It was the first {\it SOHO}-discovered object conclusively shown to be periodic \citep{hoenig06} and has now been observed on five apparitions: 1999, 2003, 2007, 2011, and 2015. 322P has an orbital period of 3.99 yr, an inclination of 12.7{\deg}, and $q=0.053$ AU (JPL Horizons). Its Tisserand parameter with respect to Jupiter of 2.3 is consistent with Jupiter-family comets (JFCs; \citealt{levison96}), although dynamical integrations to explore its prior orbital evolution are inconclusive due to its proximity to the 3:1 resonance with Jupiter \citep{hoenig06}. 

322P does not display a coma or tail in {\it SOHO} images, but its asymmetric lightcurve has repeated nearly identically each apparition \citep{lamy13} and implies a large, unresolved cross-section of dust in the photometric aperture. Its brightness has not changed substantially from apparition to apparition, implying that the depth of material lost each apparition is negligible in comparison to its total size. This suggests that 322P is significantly larger than comparably bright Kreutz comets seen at similar distances which are always destroyed prior to perihelion and are therefore estimated to be $\lesssim$10 m (e.g., \citealt{sekanina03,knight10d}). Thus, we concluded that 322P was likely $\gtrsim$100 m in diameter and potentially recoverable when far from the Sun.

322P's orbit was sufficiently constrained to attempt observations from Earth prior to its 2015 perihelion passage despite having very large positional uncertainties. Since no short period objects discovered by {\it SOHO} had ever been observed beyond the near-Sun region, such observations would be unique and highly valuable for helping to understand the population. As described below, we successfully recovered 322P and acquired optical and IR follow up observations over two months to characterize its properties and investigate its likely origin.

\section{OBSERVATIONS AND REDUCTIONS}
\label{sec:observations}
We recovered 322P with Very Large Telescope (VLT) and FORS2 imager on 2015 May 22 and obtained subsequent snapshot observations with VLT on June 16 and July 12, with Lowell Observatory's Discovery Channel Telescope (DCT) and LMI camera on June 17, 18, and 20, and with {\it Spitzer Space Telescope} on July 24. The {\it Spitzer} images were taken with InfraRed Array Camera (IRAC) through the 3.6 and 4.5~\micron{} filters \citep{werner04, fazio04}. The vast majority of all ground-based images were obtained through the $r'$ filter in order to obtain a lightcurve and look for faint dust features, but sets of $g'$ and $i'$ were also obtained May 22 and June 16 for color information. All observations were conducted at the comet's ephemeris rate. Observing circumstances and telescope details are given in Table~\ref{t:obs_circ}. We attempted to observe 322P with DCT contemporaneously with {\it Spitzer} on July 24 but could only set a weakly restrictive brightness upper limit because of clouds. Due to 322P's high southern declination, all DCT observations were obtained for short duration at high airmass and were therefore substantially noisier than the VLT observations.

\renewcommand{\baselinestretch}{0.6}
\renewcommand{\arraystretch}{1.5}

% TABLE 1
% Summary of observations
\begin{deluxetable*}{lccccccccc}  % <--- column justification (center/left/right)
\tabletypesize{\scriptsize}
\tablecolumns{10}
\tablewidth{0pt} 
\setlength{\tabcolsep}{0.05in}
\tablecaption{Summary of observations\tablenotemark{a}}
\tablehead{   % column headings
  \colhead{UT}&
  \colhead{UT}&
  \colhead{Tel.\tablenotemark{b}}&
  \colhead{r$_H$}&
  \colhead{$\Delta$}&
  \colhead{Phase}&
  \colhead{$<m_{r'}{\pm}{\sigma}_{m_{r'}}>$}&
  \colhead{$g'-r'$}&
  \colhead{$r'-i'$}&
  \colhead{Conditions}\\
  \colhead{Date}&
  \colhead{Time}&
  \colhead{}&
  \colhead{(AU)}&
  \colhead{(AU)}&
  \colhead{({\deg})}&
  \colhead{}&
  \colhead{(mag)}&
  \colhead{(mag)}&
  \colhead{}
}
\startdata
2015 May 22&4:25--7:49&VLT&2.144&1.168&$\phantom{0}$9.7&22.48$\pm$0.10&0.50$\pm$0.05&0.02$\pm$0.07&Photometric\\
2015 June 16&1:20--2:45&VLT&1.818&0.973&24.5&22.50$\pm$0.09&0.54$\pm$0.03&0.04$\pm$0.04&Photometric\\
2015 June 17&4:17--4:35&DCT&1.802&0.970&25.5&22.34$\pm$0.10&--&--&Smoke, light cirrus\\
2015 June 18&4:21--4:39&DCT&1.788&0.967&26.3&22.54$\pm$0.15&--&--&Heavy cirrus\\
2015 June 20&4:15--4:50&DCT&1.759&0.963&28.1&22.61$\pm$0.12&--&--&Possible smoke\\
2015 July 12&0:30--1:24&VLT&1.419&0.974&45.8&22.50$\pm$0.09&--&--&Thin cirrus\\
2015 July 24&3:19--3:44&DCT&1.204&0.992&54.1&$>$19.5&--&--&Clouds\\
2015 July 24&6:19--6:59&Spitzer&1.201&0.714&57.0&--&--&--&N/A\\
\enddata
\tablenotetext{a} {Geometry given for midpoint of observations.}
\tablenotetext{b} {Telescope used: VLT = Very Large Telescope UT1 (8.2 m diameter aperture, $6{\farcm}8{\times}6{\farcm}8$ field of view, $0{\farcs}25$ pixels); DCT = Discovery Channel Telescope (4.3 m, $12{\farcm}3{\times}12{\farcm}3$, $0{\farcs}24$); Spitzer = {\it Spitzer Space Telescope} (0.85 m, $5{\farcm}12{\times}5{\farcm}12$, $1{\farcs}2$).}
\label{t:obs_circ}
\label{lasttable}
\end{deluxetable*}

\renewcommand{\baselinestretch}{0.9}
\renewcommand{\arraystretch}{1.0}

The ground-based data were reduced using IRAF and IDL, following standard reduction procedures for bias removal and flat-fielding. The data were calibrated using standard stars from \citet{smith02}. The standard fields were always within $\sim$5$^\circ$ of the comet field and were taken immediately before and/or after the comet sets. This provided acceptably accurate calibrations even on non-photometric nights. We measured photometry in a series of circular apertures centered on the nucleus. Comparison star photometry was measured only on images with short enough exposure times that the stars were not trailed substantially, and the same size aperture was used for the comet and the stars to minimize the effects of variable seeing. The aperture size used for our photometry varied by night and telescope, but was chosen to maximize signal-to-noise while minimizing contamination by background objects (322P was against the Milky Way during our May observations). Typical aperture radii were 1.2--1.5 arcsec, e.g., 1.5--2$\times$ the seeing.

The {\it Spitzer} images were processed with IRAC pipeline version S19.1.0 \citep{irac}.  The 11 dithered images for each filter were combined into the comet's rest frame with the MOPEX software \citep{makovoz05}.  We measured photometry in a 6.1 arcsec radius aperture centered on the target and included an aperture correction of 1.06.  Before color correction (treated in Section \ref{sec:results}), the flux densities were 0.0102$\pm$0.0014 and 0.0389$\pm$0.0020 mJy in the 3.6 and 4.5~\micron{} filters, respectively.  Quoted uncertainties exclude the instrument's $\sim$3\% absolute calibration uncertainty.

\section{RESULTS AND DISCUSSION}\label{sec:results}
\begin{figure*}[ht]
  \centering
  \includegraphics[width=176mm]{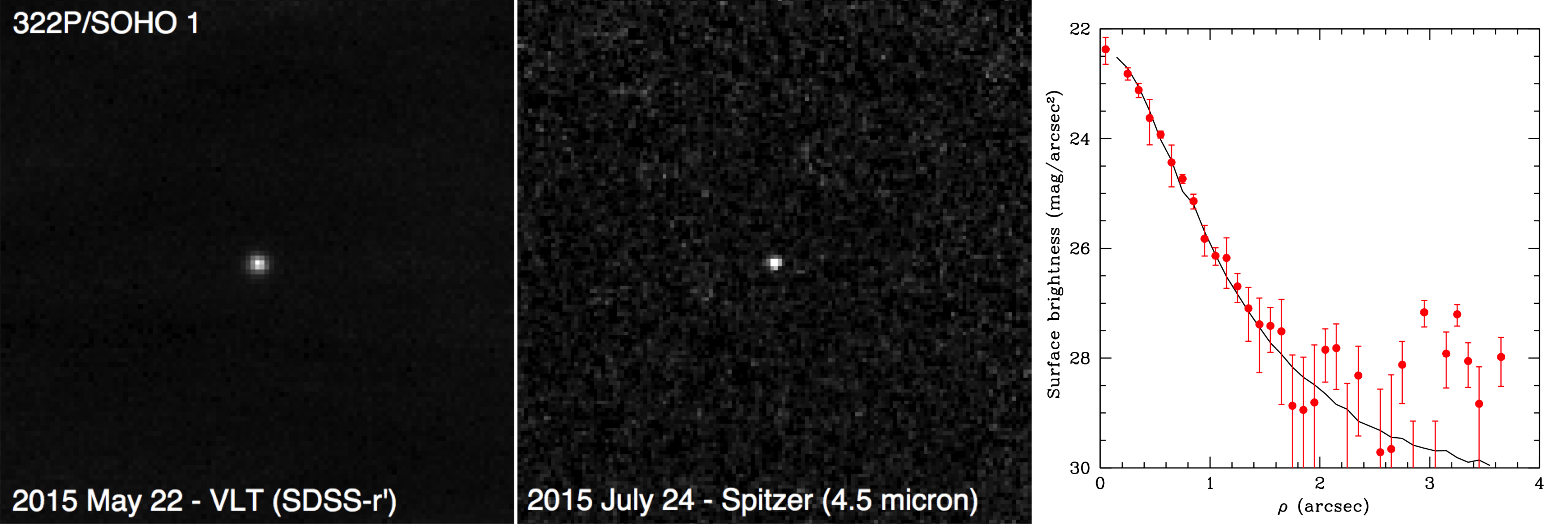}
  \caption[Images]{Stacked images from VLT (left) and {\it Spitzer} (middle) showed no evidence of cometary activity. A comparison of the comet radial profile and a stellar profile on July 12 (right) confirms there is no coma brighter than $\sim$28 mag arcsec$^{-2}$.}
  \label{fig:image}
\end{figure*}

322P did not exhibit evidence for cometary activity such as a coma, tail, or dust trail during any observations, including in nightly stacked images combining all exposures in a given filter (Figure~\ref{fig:image}). The mean magnitude each night was near $m_{r'}=22.5$ throughout the observations and is given in Table~\ref{t:obs_circ}. The standard deviation in each night's mean magnitude is considerably larger than the instrumental uncertainties ($\lesssim$0.05 mag) because they sample large portions of the lightcurve (discussed below) and because conditions were variable on non-photometric nights. Under the assumption that we observed a bare nucleus whose apparent brightness varied only due to the viewing geometry, we find a linear phase angle dependence of 0.031$\pm$0.004 mag deg$^{-1}$ and an absolute $r'$ magnitude of $20.29{\pm}0.14$. This phase function slope is close to that typically assumed for cometary nuclei, but slightly shallower than the observed mean value of 0.053 mag deg$^{-1}$ \citep{snodgrass11}. An $H, G{_1}, G{_2}$ fit \citep{muinonen10} using the online calculator from the University of Helsinki\footnote{http://www.helsinki.fi/project/psr/HG1G2/}, recommends using a single fit model for a C-type asteroid and yields $H_{r'}(C)=20.03$, although any opposition surge is not constrained by our observations (phase angle $ > 9^\circ$).

% FIGURE 1: stacked image and radial profile

\begin{figure}[t]
  \centering
  \includegraphics[width=88mm]{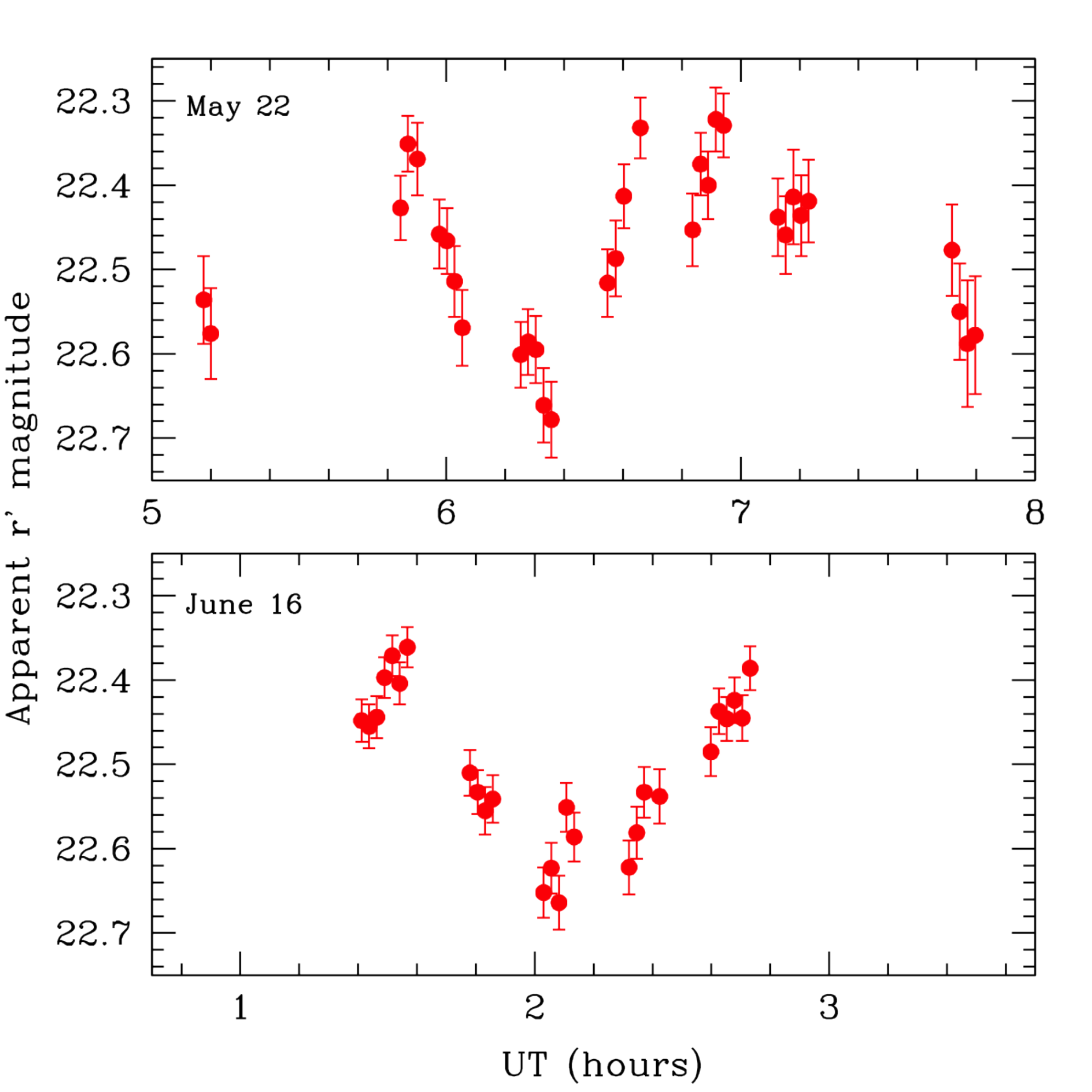}
  \caption[Lightcurves]{$r'$ lightcurves on May 22 (top) and June 16 (bottom). Both suggest a double-peaked period near $\sim$2.8 hr.}
  \label{fig:lightcurve}
\end{figure}

Data were sufficient to measure a lightcurve on only two nights, May 22 and June 16 (Figure~\ref{fig:lightcurve}). The May 22 lightcurve spanned $\sim$2.5 hr and suggests a double-peaked lightcurve that has a period of 2.5--3.0 hr and a peak-to-trough amplitude of $\sim$0.35 mag. The June 16 lightcurve covers $\sim$1.4 hr and appears to have just completed one full sinusoidal cycle. This implies a double-peaked lightcurve of $\sim$2.8 hr with a peak-to-trough amplitude of $\sim$0.3 mag. While both lightcurves are relatively noisy, they are consistent with each other and suggest a bare nucleus rotating with a period of $2.8{\pm}0.3$ hr. This rotation period is the shortest for any known comet. The peak-to-trough amplitude of $\gtrsim$0.3 mag indicates the ratio of the long to short axis is at least $10^{0.4{\Delta}m} = 1.3:1$. The density for a strengthless body implied by the combination of the rotation period and axial ratio is $>$1000~kg~m$^{-3}$. This is significantly higher than the limits found this way for other comets ($\sim$600~kg~m$^{-3}$; \citealt{snodgrass06}), but compatible with typical asteroids ($\sim$2200~kg~m$^{-3}$; \citealt{pravec02}). Spacecraft results have confirmed the low density of cometary nuclei (e.g., 533~kg~m$^{-3}$ for 67P; \citealt{patzold16}).

% FIGURE 2: Lightcurves

\begin{figure}[t]
  \centering
  \includegraphics[width=88mm]{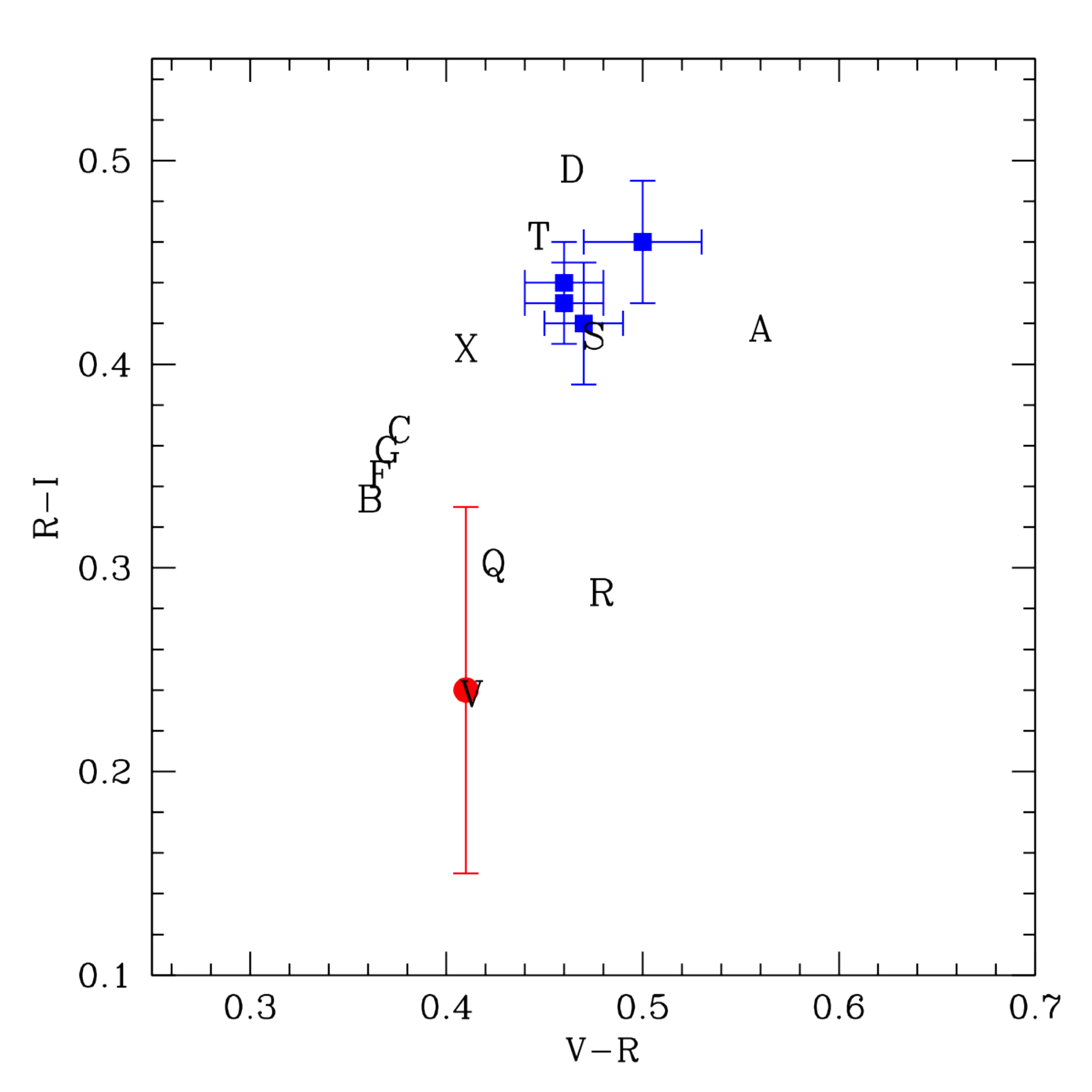}
  \caption[Nucleus properties]{Average colors for 322P (red circle) compared with asteroids (letters designating class; \citealt{dandy03}), and comets (blue squares; \citealt{jewitt15} and references therein). 322P's SDSS colors were translated to Johnson colors following \citet{jordi06} for comparison with the other datasets.}
  \label{fig:nuc_prop}
\end{figure}

We measured $g'-r'$ and $r'-i'$ colors on May 22 and June 16 by averaging $r'$ images acquired either side of the $g'$ or $i'$ filter images to remove any lightcurve effects. We find $g'-r'=0.52{\pm}0.04$ and $r'-i'=0.03{\pm}0.06$; the average and standard deviation of all color measurements for each night are given in Table~\ref{t:obs_circ}. There was no clear evidence for color variations as a function of rotational phase, although we note that the data are noisy. We converted these to $V-R=0.41{\pm}0.04$ and $R-I=0.24{\pm}0.09$ using the translations from \citet{jordi06} for ease of comparison with existing datasets. The $V-R$ color is consistent with the colors found for active and inactive JFCs and long period comets, but bluer than these comets in $R-I$ (\citealt{jewitt15} and references therein; Figure~\ref{fig:nuc_prop}). When compared with asteroids, 322P's color best matches V-types and is also consistent with Q-types. A more definitive analysis of the colors as compared to other small bodies would require improved uncertainties in our color measurements and/or measurements at additional wavelengths. 

% FIGURE 3: Asteroidal colors

\begin{figure}[t]
  \centering
  \includegraphics[width=88mm]{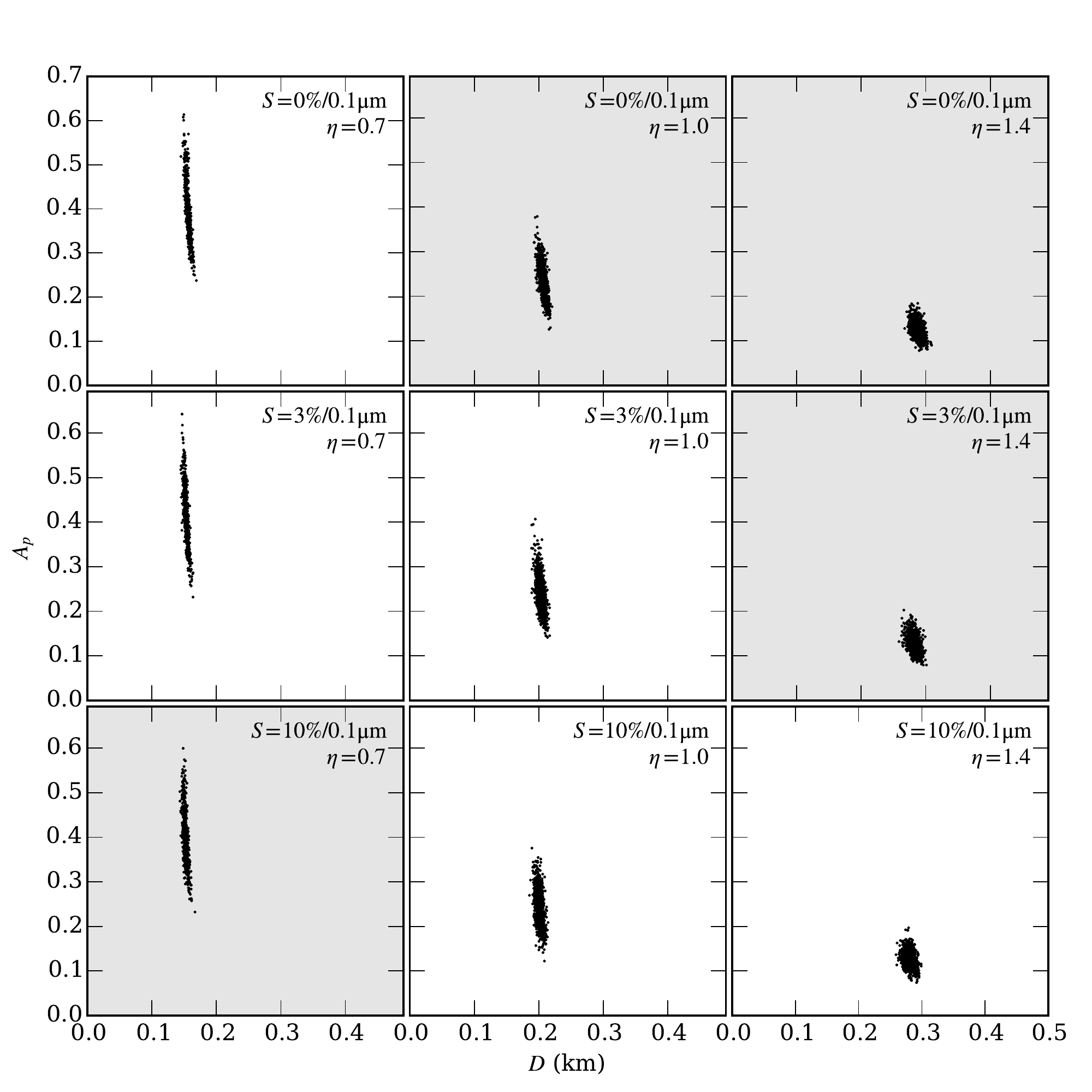}
  \caption[NEATM fit]{Near-Earth asteroid thermal model best-fit diameter $D$ and geometric albedo $A_p$ pairs from our Monte Carlo analysis of the optical and IR data.  A subset of the spectral slope $S$, and beaming parameter $\eta$ values are presented here to illustrate their minimal effects on the retrieved values. The gray-shaded panels have higher $\chi^2$ values suggesting those $S$, $\eta$ pairs are less likely.}
  \label{fig:model}
  \label{lastfig}
\end{figure}

The combination of optical and IR observations allows us to constrain 322P's albedo and size with Near Earth Asteroid Thermal Modeling (NEATM; \citealt{harris98}). We modeled a range of optical-near-IR spectral slopes, $S$ = $-$1 to $+$10 \% per 0.1 $\mu$m, roughly corresponding to the O, C, S, and D-type asteroid spectral archetypes of \citet{demeo09}, and IR beaming parameters, ${\eta} = 0.7-3.0$.  The neutral to moderately red slopes, $0-3$\% per 0.1~$\mu$m, are potentially the most consistent with the suggested Q or V spectral types.  The chosen range of $\eta$ is based on the distribution of beaming parameters in WISE observations of near-Earth objects \citep{mainzer11}. For each combination of $S$ and $\eta$, we let diameter, $D$, and geometric albedo, $A_p$, vary freely. The model takes the V-band absolute magnitude (converted from $r'$ using the translations from \citealt{jordi06}), the phase function, and the IRAC photometry as input parameters. For each best-fit model, we computed color corrections for the broad IRAC bandpasses \citep{irac} and re-executed the fitting procedure. Because we have three independent parameters (phase function and absolute magnitude are anti-correlated) and four unknown parameters, we explored parameter space through a Monte Carlo simulation. We use the $\chi^2$ statistic to assess relative goodness of fit, but realize its absolute value has little meaning in this context. For each $S$ and $\eta$ combination, we repeated our fitting procedure with 1000 new input data sets normally distributed about the observed values using their estimated uncertainties.  Figure~\ref{fig:model} shows example best-fit $D$ and $A_p$ pairs.  We find a weak dependence of $D$ and $A_p$ on chosen $S$ and $\eta$ values.  However, some sets of $S$ and $\eta$ produce higher mean $\chi^2$ values. Considering the combinations of $S$ and $\eta$ that yield the most fits with $\chi^2\lesssim1.0$, 322P's diameter is likely 150--320~m, with albedo between 0.09 and 0.42 (albedo and diameter are inversely correlated).

% FIGURE 4: Diameter and albedo retrieval

The large uncertainties on our optical magnitudes combined with the number of free parameters prevents unequivocally deriving a unique solution for 322P's diameter and geometric albedo. Despite this, 322P's geometric albedo is evidently substantially higher than typical values for cometary nuclei (0.02--0.06; \citealt{lamy04}). Such low geometric albedos are incompatible with the {\it Spitzer} 3.6 $\mu$m data and would require an unusually high $\eta$. However, $A_p$ is consistent with the geometric albedos of many asteroids, notably including V- and Q-types \citep{thomas11} and those having small-$q$ \citep{campins09}.

The modeling confirms our assumption that 322P was substantially larger than typical Kreutz comets seen close to the Sun. Other than the Kreutz family and fragments of recently split comets like C/1999 S4 LINEAR \citep{weaver01}, 322P is the smallest comet nucleus ever identified \citep{snodgrass11}. Such small comets may not be necessarily rare but simply difficult to detect because they are so faint. 322P is not unusual compared to the sizes of known NEAs, where $\sim$100 m diameter objects are common \citep{mainzer14}

As shown in Figure~\ref{fig:image}, the radial profile was consistent with profiles of untrailed stars suggesting cometary activity is minimal or absent. We estimate that any undetected coma had a surface brightness $\sim$28 mag arcsec$^{-2}$ or fainter at a distance of 2.5 arcsec on June 16. Assuming a steady state coma, we estimate the total magnitude of any undetected coma to be $>$24.0 following \citet{jewitt84}, and convert this to $Af{\rho} < 0.03$ cm \citep{ahearn84} for an assumed dust albedo of 0.04. 

We next attempt to constrain 322P's lifetime with some order of magnitude calculations about its total mass loss per orbit. The undetected coma upper limit can be converted to a cross section of dust and then to a combined dust mass of $\lesssim$200 kg using standard assumptions about the dust's geometric albedo (0.04), density (3000 kg m$^{-3}$), and radius (1 $\mu$m). If the albedo or density are closer to the values derived for the nucleus above, the dust mass will be even smaller. Assuming the dust crosses our photometric aperture at $\lesssim$1 km s$^{-1}$ sets a conservative upper limit to the mass loss rate of ${\sim}10^{-4}$ kg s$^{-1}$ at 1.419 AU. We repeat the calculation when 322P is in the {\it SOHO} field of view, where it reaches a peak $m_V{\sim}6$ in an aperture of radius ${\sim}145\arcsec$ \citep{lamy13} at perihelion. Near perihelion, the sublimation lifetime of dust grains ($\sim$100 sec for amorphous olivine; \citealt{kimura02}) may dominate the aperture crossing time ($\gtrsim$10$^4$ sec), so we estimate the dust spends $\gtrsim$100--10$^4$ sec in the aperture, yielding a peak mass loss rate of $\lesssim$20--2000 kg s$^{-1}$. Assuming a power-law slope between these points and integrating around the orbit, we find that the total mass lost is dominated by activity at perihelion and is highly assumption dependent. For the low and high activity cases, the total mass lost per orbit is ${\lesssim}2{\times}10^{6}-2{\times}10^{8}$ kg. For a diameter of 200 m and a density of 1500 kg m$^{-3}$, 322P's active lifetime is therefore likely to be $10^3-10^5$ yr, which is comparable to or shorter than estimates of the total dynamical lifetimes of JFCs and NEAs, $10^{5}-10^{6}$ yr (e.g., \citealt{bailey92b,levison94}). Thus, 322P may be destroyed by sublimation-driven mass loss sooner than it is removed by dynamical processes, although we caution that the activity calculations are extremely uncertain.

\section{CONCLUSIONS}
Our observations of 322P/SOHO 1 suggest that it may be of classically asteroidal rather than cometary origin. It appears inactive, has a rotation period that is faster than any known comet and implies a density $>$1000 kg m$^{-3}$, has a higher albedo than has ever been measured for a comet nucleus, and has colors rather atypical for a comet and most consistent with V- and Q-type asteroids. The unweathered spectra of Q-types seem to correlate with recent close approaches to planets \citep{binzel10}; given 322P's small $q$, similar processing may be occurring. However, no comet nucleus has ever been studied following such a close perihelion passage, and we cannot exclude the possibility that 322P's unusual properties are caused by its extreme orbit, where equilibrium temperatures exceed 1000 K at perihelion. It is possible that 322P's repeated close approaches to the Sun devolatilized its outer layers and caused changes to the surface, such as annealing, that are sufficient to alter its properties (albedo, color) and increase its strength against rotational breakup.

If 322P is an ``active asteroid,'' it has the smallest $q$ of any known asteroid. All asteroids with $q<0.15$ AU reach perihelion within {\it SOHO}'s field of view, but we are only aware of Phaethon having been detected. {\it SOHO} data are routinely scoured for moving objects by multiple comet hunters, so it is unlikely that all such objects would have gone undetected if visible. Since both Phaethon and 322P display evidence of being active inside $\sim$0.15 AU, the lack of detections of any other small-$q$ asteroids in the {\it SOHO} fields of view is likely significant. All of the small-$q$ asteroids studied by \citet{jewitt13} had absolute magnitudes brighter than 322P, so if they are active by the same mechanism as 322P, they should have been detectable when in the {\it SOHO} field of view despite their larger $q$ values. Thus, 322P may be different from typical small-$q$ asteroids.

One possible difference is that 322P may have undergone a recent breakup that exposed parts of its interior which could be more easily lost on subsequent perihelion passages. Three other {\it SOHO}-discovered objects have been suggested to be dynamically linked to 322P as part of the ``Kracht-2 group'' \citep{iauc8983} and could be fragments of such an event. Since 322P's orbit is well beyond the distances where tidal forces should be significant (\citealt{knight13b} and references therein), breakup would likely have been caused by other mechanisms such as rotational spin-up or heating-driven fracture. Such processes have recently been suggested by \citet{granvik16} as the mechanisms by which small-$q$ asteroids are destroyed. Thus, 322P may be providing a near-real time glimpse at a common end state of small bodies in the solar system and warrants further inquiry.

\section*{ACKNOWLEDGMENTS}
We thank Mikael Granvik for a prompt and helpful review. We also thank our observatory support staff: Lorena Faundez, Linda Schmidtobreick, Henri Boffin, Patricia Guajardo, Joseph Anderson and Sergio Vera Urrutia at VLT, and Heidi Larson and Jason Sanborn at DCT, as well as Joe Llama for collecting the July 24 DCT data. We are grateful to Karl Battams, Jon Giorgini, and Gareth Williams for help in refining 322P's positional uncertainty prior to our observations. M.M.K. was supported by NASA Planetary Astronomy Program grant NNX14AG81G. A.F. was supported by UK STFC grant ST/L000709/1. C.S. was supported by a UK STFC Ernest Rutherford Fellowship. 

These results made use of the Discovery Channel Telescope at Lowell Observatory. Lowell is a private, non-profit institution dedicated to astrophysical research and public appreciation of astronomy and operates the DCT in partnership with Boston University, the University of Maryland, the University of Toledo, Northern Arizona University, and Yale University.  LMI construction was supported by grant AST-1005313 from the National Science Foundation.

This work is based in part on observations made with the {\it Spitzer Space Telescope}, which is operated by the Jet Propulsion Laboratory, California Institute of Technology under a contract with NASA.


\begin{thebibliography}{}
\expandafter\ifx\csname natexlab\endcsname\relax\def\natexlab#1{#1}\fi

\bibitem[{{A'Hearn} {et~al.}(1984){A'Hearn}, {Schleicher}, {Millis}, {Feldman},
  \& {Thompson}}]{ahearn84}
{A'Hearn}, M.~F., {Schleicher}, D.~G., {Millis}, R.~L., {Feldman}, P.~D., \&
  {Thompson}, D.~T. 1984, \aj, 89, 579

\bibitem[{{Bailey} {et~al.}(1992){Bailey}, {Chambers}, \& {Hahn}}]{bailey92b}
{Bailey}, M.~E., {Chambers}, J.~E., \& {Hahn}, G. 1992, \aap, 257, 315

\bibitem[{{Biesecker} {et~al.}(2002){Biesecker}, {Lamy}, {St.~Cyr}, {Llebaria},
  \& {Howard}}]{biesecker02}
{Biesecker}, D.~A., {Lamy}, P., {St.~Cyr}, O.~C., {Llebaria}, A., \& {Howard},
  R.~A. 2002, Icarus, 157, 323

\bibitem[{{Binzel} {et~al.}(2010){Binzel}, {Morbidelli}, {Merouane}, {DeMeo},
  {Birlan}, {Vernazza}, {Thomas}, {Rivkin}, {Bus}, \& {Tokunaga}}]{binzel10}
{Binzel}, R.~P., {Morbidelli}, A., {Merouane}, S., {et~al.} 2010, \nat, 463,
  331

\bibitem[{{Bottke} {et~al.}(2002){Bottke}, {Morbidelli}, {Jedicke}, {Petit},
  {Levison}, {Michel}, \& {Metcalfe}}]{bottke02}
{Bottke}, W.~F., {Morbidelli}, A., {Jedicke}, R., {et~al.} 2002, Icarus, 156,
  399

\bibitem[{{Campins} {et~al.}(2009){Campins}, {Kelley}, {Fern{\'a}ndez},
  {Licandro}, \& {Hargrove}}]{campins09}
{Campins}, H., {Kelley}, M.~S., {Fern{\'a}ndez}, Y., {Licandro}, J., \&
  {Hargrove}, K. 2009, Earth Moon and Planets, 105, 159

\bibitem[{{Dandy} {et~al.}(2003){Dandy}, {Fitzsimmons}, \&
  {Collander-Brown}}]{dandy03}
{Dandy}, C.~L., {Fitzsimmons}, A., \& {Collander-Brown}, S.~J. 2003, \icarus,
  163, 363

\bibitem[{{DeMeo} {et~al.}(2009){DeMeo}, {Binzel}, {Slivan}, \&
  {Bus}}]{demeo09}
{DeMeo}, F.~E., {Binzel}, R.~P., {Slivan}, S.~M., \& {Bus}, S.~J. 2009, Icarus,
  202, 160

\bibitem[{{Farinella} {et~al.}(1994){Farinella}, {Froeschle}, {Gonczi}, {Hahn},
  {Morbidelli}, \& {Valsecchi}}]{farinella94}
{Farinella}, P., {Froeschle}, C., {Gonczi}, R., {et~al.} 1994, \nat, 371, 314

\bibitem[{{Fazio} {et~al.}(2004){Fazio}, {Hora}, {Allen}, {Ashby}, {Barmby},
  {Deutsch}, {Huang}, {Kleiner}, {Marengo}, {Megeath}, {Melnick}, {Pahre},
  {Patten}, {Polizotti}, {Smith}, {Taylor}, {Wang}, {Willner}, {Hoffmann},
  {Pipher}, {Forrest}, {McMurty}, {McCreight}, {McKelvey}, {McMurray}, {Koch},
  {Moseley}, {Arendt}, {Mentzell}, {Marx}, {Losch}, {Mayman}, {Eichhorn},
  {Krebs}, {Jhabvala}, {Gezari}, {Fixsen}, {Flores}, {Shakoorzadeh}, {Jungo},
  {Hakun}, {Workman}, {Karpati}, {Kichak}, {Whitley}, {Mann}, {Tollestrup},
  {Eisenhardt}, {Stern}, {Gorjian}, {Bhattacharya}, {Carey}, {Nelson},
  {Glaccum}, {Lacy}, {Lowrance}, {Laine}, {Reach}, {Stauffer}, {Surace},
  {Wilson}, {Wright}, {Hoffman}, {Domingo}, \& {Cohen}}]{fazio04}
{Fazio}, G.~G., {Hora}, J.~L., {Allen}, L.~E., {et~al.} 2004, \apjs, 154, 10

\bibitem[{{Gladman} {et~al.}(1997){Gladman}, {Migliorini}, {Morbidelli},
  {Zappala}, {Michel}, {Cellino}, {Froeschle}, {Levison}, {Bailey}, \&
  {Duncan}}]{gladman97}
{Gladman}, B.~J., {Migliorini}, F., {Morbidelli}, A., {et~al.} 1997, Science,
  277, 197

\bibitem[{{Granvik} {et~al.}(2016){Granvik}, {Morbidelli}, {Jedicke}, {Bolin},
  {Bottke}, {Beshore}, {Vokrouhlick{\'y}}, {Delb{\`o}}, \&
  {Michel}}]{granvik16}
{Granvik}, M., {Morbidelli}, A., {Jedicke}, R., {et~al.} 2016, \nat, 530, 303

\bibitem[{{Greenstreet} {et~al.}(2012){Greenstreet}, {Ngo}, \&
  {Gladman}}]{greenstreet12}
{Greenstreet}, S., {Ngo}, H., \& {Gladman}, B. 2012, \icarus, 217, 355

\bibitem[{{Harris}(1998)}]{harris98}
{Harris}, A.~W. 1998, \icarus, 131, 291

\bibitem[{{H{\"o}nig}(2006)}]{hoenig06}
{H{\"o}nig}, S.~F. 2006, \aap, 445, 759

\bibitem[{{Jewitt}(2013)}]{jewitt13}
{Jewitt}, D. 2013, \aj, 145, 133

\bibitem[{{Jewitt}(2015)}]{jewitt15}
---. 2015, \aj, 150, 201

\bibitem[{{Jewitt} \& {Danielson}(1984)}]{jewitt84}
{Jewitt}, D., \& {Danielson}, G.~E. 1984, \icarus, 60, 435

\bibitem[{{Jewitt} \& {Li}(2010)}]{jewitt10}
{Jewitt}, D., \& {Li}, J. 2010, \aj, 140, 1519

\bibitem[{{Jordi} {et~al.}(2006){Jordi}, {Grebel}, \& {Ammon}}]{jordi06}
{Jordi}, K., {Grebel}, E.~K., \& {Ammon}, K. 2006, \aap, 460, 339

\bibitem[{{Kimura} {et~al.}(2002){Kimura}, {Mann}, {Biesecker}, \&
  {Jessberger}}]{kimura02}
{Kimura}, H., {Mann}, I., {Biesecker}, D.~A., \& {Jessberger}, E.~K. 2002,
  Icarus, 159, 529

\bibitem[{{Knight} {et~al.}(2010){Knight}, {A'Hearn}, {Biesecker}, {Faury},
  {Hamilton}, {Lamy}, \& {Llebaria}}]{knight10d}
{Knight}, M.~M., {A'Hearn}, M.~F., {Biesecker}, D.~A., {et~al.} 2010, \aj, 139,
  926

\bibitem[{{Knight} \& {Walsh}(2013)}]{knight13b}
{Knight}, M.~M., \& {Walsh}, K.~J. 2013, \apjl, 776, L5

\bibitem[{{Kracht} {et~al.}(2008){Kracht}, {Marsden}, {Battams}, \&
  {Sekanina}}]{iauc8983}
{Kracht}, R., {Marsden}, B.~G., {Battams}, K., \& {Sekanina}, Z. 2008,
  \iaucirc, 8983

\bibitem[{{Laine}(2015)}]{irac}
{Laine}, S., ed. 2015, IRAC Instrument Handbook (Pasadena: {Spitzer Science
  Center})

\bibitem[{{Lamy} {et~al.}(2013){Lamy}, {Faury}, {Llebaria}, {Knight},
  {A'Hearn}, \& {Battams}}]{lamy13}
{Lamy}, P., {Faury}, G., {Llebaria}, A., {et~al.} 2013, \icarus, 226, 1350

\bibitem[{{Lamy} {et~al.}(2004){Lamy}, {Toth}, {Fernandez}, \&
  {Weaver}}]{lamy04}
{Lamy}, P.~L., {Toth}, I., {Fernandez}, Y.~R., \& {Weaver}, H.~A. 2004, in
  Comets II, ed. M.~C. {Festou}, H.~U. {Keller}, \& H.~A. {Weaver} (Univ. of
  Arizona Press/Lunar Planet. Inst., Tucson, AZ/Houston, TX), 223--264

\bibitem[{{Levison}(1996)}]{levison96}
{Levison}, H.~F. 1996, in Astronomical Society of the Pacific Conference
  Series, Vol. 107, Completing the Inventory of the Solar System, ed.
  T.~{Rettig} \& J.~M. {Hahn} (San Francisco, CA: ASP), 173--191

\bibitem[{{Levison} \& {Duncan}(1994)}]{levison94}
{Levison}, H.~F., \& {Duncan}, M.~J. 1994, \icarus, 108, 18

\bibitem[{{Li} \& {Jewitt}(2013)}]{li13b}
{Li}, J., \& {Jewitt}, D. 2013, \aj, 145, 9pp

\bibitem[{{Mainzer} {et~al.}(2011){Mainzer}, {Bauer}, {Grav}, {Masiero},
  {Cutri}, {Dailey}, {Eisenhardt}, {McMillan}, {Wright}, {Walker}, {Jedicke},
  {Spahr}, {Tholen}, {Alles}, {Beck}, {Brandenburg}, {Conrow}, {Evans},
  {Fowler}, {Jarrett}, {Marsh}, {Masci}, {McCallon}, {Wheelock}, {Wittman},
  {Wyatt}, {DeBaun}, {Elliott}, {Elsbury}, {Gautier}, {Gomillion}, {Leisawitz},
  {Maleszewski}, {Micheli}, \& {Wilkins}}]{mainzer11}
{Mainzer}, A., {Bauer}, J., {Grav}, T., {et~al.} 2011, \apj, 731, 53

\bibitem[{{Mainzer} {et~al.}(2014){Mainzer}, {Bauer}, {Grav}, {Masiero},
  {Cutri}, {Wright}, {Nugent}, {Stevenson}, {Clyne}, {Cukrov}, \&
  {Masci}}]{mainzer14}
---. 2014, \apj, 784, 110

\bibitem[{{Makovoz} \& {Khan}(2005)}]{makovoz05}
{Makovoz}, D., \& {Khan}, I. 2005, in Astronomical Society of the Pacific
  Conference Series, Vol. 347, Astronomical Data Analysis Software and Systems
  XIV, ed. {P.~Shopbell, M.~Britton, \& R.~Ebert} (San Francisco, CA: ASP), 81

\bibitem[{{Muinonen} {et~al.}(2010){Muinonen}, {Belskaya}, {Cellino},
  {Delb{\`o}}, {Levasseur-Regourd}, {Penttil{\"a}}, \& {Tedesco}}]{muinonen10}
{Muinonen}, K., {Belskaya}, I.~N., {Cellino}, A., {et~al.} 2010, \icarus, 209,
  542

\bibitem[{{P{\"a}tzold} {et~al.}(2016){P{\"a}tzold}, {Andert}, {Hahn}, {Asmar},
  {Barriot}, {Bird}, {H{\"a}usler}, {Peter}, {Tellmann}, {Gr{\"u}n},
  {Weissman}, {Sierks}, {Jorda}, {Gaskell}, {Preusker}, \&
  {Scholten}}]{patzold16}
{P{\"a}tzold}, M., {Andert}, T., {Hahn}, M., {et~al.} 2016, \nat, 530, 63

\bibitem[{{Pravec} {et~al.}(2002){Pravec}, {Harris}, \&
  {Michalowski}}]{pravec02}
{Pravec}, P., {Harris}, A.~W., \& {Michalowski}, T. 2002, in Asteroids III, ed.
  W.~F. {Bottke}, Jr., A.~{Cellino}, P.~{Paolicchi}, \& R.~P. {Binzel}
  (University of Arizona Press, Tucson), 113--122

\bibitem[{{Sekanina}(2003)}]{sekanina03}
{Sekanina}, Z. 2003, \apj, 597, 1237

\bibitem[{{Smith} {et~al.}(2002){Smith}, {Tucker}, {Allam}, \&
  {Jorgensen}}]{smith02}
{Smith}, J.~A., {Tucker}, D.~L., {Allam}, S.~S., \& {Jorgensen}, A.~M. 2002, in
  Bulletin of the American Astronomical Society, Vol.~34, American Astronomical
  Society Meeting Abstracts, 1272

\bibitem[{{Snodgrass} {et~al.}(2011){Snodgrass}, {Fitzsimmons}, {Lowry}, \&
  {Weissman}}]{snodgrass11}
{Snodgrass}, C., {Fitzsimmons}, A., {Lowry}, S.~C., \& {Weissman}, P. 2011,
  \mnras, 414, 458

\bibitem[{{Snodgrass} {et~al.}(2006){Snodgrass}, {Lowry}, \&
  {Fitzsimmons}}]{snodgrass06}
{Snodgrass}, C., {Lowry}, S.~C., \& {Fitzsimmons}, A. 2006, \mnras, 373, 1590

\bibitem[{{Thomas} {et~al.}(2011){Thomas}, {Trilling}, {Emery}, {Mueller},
  {Hora}, {Benner}, {Bhattacharya}, {Bottke}, {Chesley}, {Delb{\'o}}, {Fazio},
  {Harris}, {Mainzer}, {Mommert}, {Morbidelli}, {Penprase}, {Smith}, {Spahr},
  \& {Stansberry}}]{thomas11}
{Thomas}, C.~A., {Trilling}, D.~E., {Emery}, J.~P., {et~al.} 2011, \aj, 142, 85

\bibitem[{{Weaver} {et~al.}(2001){Weaver}, {Sekanina}, {Toth}, {Delahodde},
  {Hainaut}, {Lamy}, {Bauer}, {A'Hearn}, {Arpigny}, {Combi}, {Davies},
  {Feldman}, {Festou}, {Hook}, {Jorda}, {Keesey}, {Lisse}, {Marsden}, {Meech},
  {Tozzi}, \& {West}}]{weaver01}
{Weaver}, H.~A., {Sekanina}, Z., {Toth}, I., {et~al.} 2001, Science, 292, 1329

\bibitem[{{Werner} {et~al.}(2004){Werner}, {Roellig}, {Low}, {Rieke}, {Rieke},
  {Hoffmann}, {Young}, {Houck}, {Brandl}, {Fazio}, {Hora}, {Gehrz}, {Helou},
  {Soifer}, {Stauffer}, {Keene}, {Eisenhardt}, {Gallagher}, {Gautier}, {Irace},
  {Lawrence}, {Simmons}, {Van Cleve}, {Jura}, {Wright}, \&
  {Cruikshank}}]{werner04}
{Werner}, M.~W., {Roellig}, T.~L., {Low}, F.~J., {et~al.} 2004, \apjs, 154, 1

\end{thebibliography}
\end{document}